\documentclass[12pt]{article}

\usepackage[latin2]{inputenc}
\usepackage[usenames]{color}
\usepackage{latexsym}
\usepackage{epsf}
\usepackage{epsfig}
\usepackage{amssymb,amsmath}
\usepackage{slashed}

\def\nn{\nonumber \\}

\def\cC{{\cal C}}

\def\cR{{\cal R}}
\def\cL{{\cal L}}

\def\Z2{{\mathbb Z}_2}

\def\pa{\partial}
\def\Ln{{\pounds_n}}

\def\bG{{\overline{G}}}
\def\bL{{\overline{\Lambda}}}

\def\bLn{\pounds_n}

\def\beq{\begin{equation}}
\def\eeq{\end{equation}}

\def\bea{\begin{eqnarray}}
\def\eea{\end{eqnarray}}

\def\bi{\begin{itemize}}
\def\ei{\end{itemize}}

\def\be{\begin{enumerate}}
\def\ee{\end{enumerate}}

\def\bc{\begin{center}}
\def\ec{\end{center}}

\def\ud{\mathrm{d}}

\newlength{\dinwidth}
\newlength{\dinmargin}
\begin{document}

\thispagestyle{empty}
\begin{flushright}
\end{flushright}

\vspace*{15mm}

\begin{center}
{\Large\bf 
Large-scale structure challenges dilaton gravity\\
in a 5D brane scenario with AdS bulk}
\vspace*{5mm}
\end{center}

\vspace*{5mm} \noindent
\vskip 0.5cm
\centerline{\bf
Dominika Konikowska
}
\vskip 5mm
\centerline{\em Fakult\"at f\"ur Physik, Universit\"at Bielefeld, Postfach 100131, 33501 Bielefeld,
Germany}
\centerline{\em Email: dkonikowska@physik.uni-bielefeld.de}

\vskip 15mm

{\bf Abstract:} 
We study a theory of dilaton gravity in a 5-dimensional brane scenario,
with a non-minimal coupling of the dilaton to the matter content of the universe localized on the brane. 
The effective gravitational equations at the brane are derived in the Einstein frame in the covariant approach, 
addressing certain misconceptions in the literature.
We then investigate whether the observed large-scale structure of the universe 
can exist on the brane in this dilaton gravity scenario with an exact anti de Sitter bulk, 
assuming that the matter energy-momentum tensor has the form of an inhomogeneous perfect fluid.
The corresponding constraint on the spatial derivative of the matter energy density is derived,
and subsequently quantified using the current limits resulting from searches for variation of the Newton's constant.
By confronting it with the observational data from galaxy surveys,
we show that up to scales of the order of $10^4\,$Mpc, the derived bound on the spatial derivative of the 
matter energy density does not allow for the existence of the large-scale structure as observed today.
Thus, such a dilaton gravity brane scenario is ruled out.

\newpage

\section{ Introduction }

General Relativity provides a successful description of gravitational interactions 
at macroscopic scales, and has passed numerous consistency and experimental tests. 
Nevertheless, it is not without shortcomings. 
For example, the resistance to quantization, or the lack of a unified description 
of gravity and the gauge interactions of the Standard Model, 
are usually considered to be crucial problems. 
Consequently, the search for an extended theory of gravity 
- with Einstein's General Relativity as the main part of its structure - continues. 

String theories (or the encompassing M-theory) 
are often regarded as the most promising proposal for a `theory of everything'.
Moreover, they provide a viable candidate for a quantum theory of gravity. 
One of the most commonly recognized approaches to addressing string theories 
in the field theory language employs the low-energy effective action \cite{lowenergy}, 
where only the massless modes, whose number is finite, are considered. 
The low-energy effective action in string theories 
is often represented in the form of the $\alpha'$ expansion
($\alpha'=M_s^{-2}$, where $M_s$ denotes the fundamental string energy scale), 
which can include terms for various particles - depending on the type of the string theory. 
Nevertheless, two fields are common for all string theories: 
the metric tensor $g_{\mu\nu}$ and the dilaton field $\phi$. 
If we restrict the low-energy effective action in string theories to gravity and the dilaton,
it will yield at the leading order the Einstein gravity coupled to the dilaton.

In the present work, we will address a theory of dilaton gravity, employed in a brane scenario 
with a general coupling of the dilaton to the matter content of the universe.
We will investigate whether the observed large-scale structure of the universe
can exist on the brane in an AdS$_5$ bulk,
assuming that the matter energy-momentum tensor has the form of an inhomogeneous perfect fluid. 

It is essential to develop a framework for addressing 
at least some of the important issues concerning extended theories of gravity as motivated by string theories.
In Ref.\ \cite{paper2} a physically viable theory of higher order dilaton gravity was established 
in higher-dimensional spacetime,
reflecting some of the basic properties of the effective low-energy action in string theories 
if restricted to gravity and the dilaton.
It should be underlined, that the terms with up to four derivatives 
turned out to be exactly the same as in the effective low-energy action derived from string theories \cite{meissner}.
Moreover, the discussion of the $O(d,d)$ symmetry conducted in Ref.\ \cite{paper2}
indicates that the effective dilaton gravity action in string theories
can include this theory of higher order dilaton gravity also at the level of more than four derivatives.
Consequently, addressing higher order dilaton gravity and its effective 4-dimensional description
will allow for studying e.g.\ the influence on cosmology of such modifications to General Relativity 
as are motivated by string theories - and their potentially observable effects. 

In order to study the phenomenological implications of such an alternative theory of gravity, 
its effective 4-dimensional description has to be established. 
In the present work, the idea of localizing the Standard Model on a brane  
embedded in a higher-dimensional spacetime \cite{brane} will be adopted. 
Since the M-theory based works by Ho\v{r}ava and Witten \cite{HoWi}, 
this type of scenario has become one of the most often employed approaches to higher-dimensional theories,
and bulk related corrections to the gravity interactions at the brane were investigated in many models \cite{review}.
For the string frame and the bulk action given by the higher order dilaton gravity theory 
constructed in Ref.\ \cite{paper2},
the derivation of the effective gravitational equations at the co-dimension 1 brane 
was addressed in detail in Ref.\ \cite{paper3} in the covariant approach.\footnote{
  For the standard gravity, the effective equations of motion at the brane  
  were derived in the covariant approach in Ref.\ \cite{SaShMa}.
  For the lowest order dilaton gravity, the effective equations at the brane were derived 
  (but not in a fully covariant way) in Ref.\ \cite{MaWa},
  whereas in Ref.\ \cite{MeBa} the covariant approach was employed for cosmological applications.}

Since dilaton gravity is a theory of gravity coupled to a scalar field, it can be formulated in various frames
related by conformal (Weyl) transformations.
It should be underlined here, 
that if a brane matter term $\cL_m$ is included into the Lagrangian in one of these frames, 
upon the conformal transformation to another frame its coefficient will change. 
In particular, if constant in one frame, it will become dilaton dependent in others. 

There is no clear consensus\footnote{
  For various approaches see e.g.\ Ref.\ \cite{conformal} and references therein.} 
in the modern literature as to which conformal frame is the natural physical frame, 
and in which frame the coupling of the scalar field to the matter content of the universe should be minimal.
Therefore in the present work we will briefly readdress 
the derivation of the effective equations of motion at the brane \cite{paper3}, 
but this time with a general non-minimal coupling $f(\phi) \, \cL_m$ of the dilaton 
to the brane matter Lagrangian in the Einstein frame, which is the usual framework employed in cosmology. 

In order to investigate basic properties of this dilaton gravity scenario 
and its effective gravitational equations at the brane, 
I will adopt two assumptions which are crucial to many models in the modern literature. 
In cosmological considerations, the matter content of the universe is usually described by a perfect fluid. 
In higher-dimensional scenarios, the bulk is often assumed to be of the anti de Sitter (AdS) type.
Its popularity is mainly due to its use in the widely studied AdS/CFT correspondence \cite{ads/cft}, 
as well as it being vital for many higher-dimensional scenarios 
(it is e.g.\ a natural solution for the bulk in the Randall-Sundrum model \cite{RS}).
Consequently, the main goal of the present work is to investigate 
whether the existence of the inhomogeneous perfect fluid on the brane 
- as would be expected to describe the observed large-scale structure of the universe -
is allowed for the AdS$_5$ bulk. 
The constraint on the spatial derivative of the matter energy density
will be derived employing only the effective gravitational brane equations
and the 4-dimensional Bianchi identity.
It will be quantified using the current limits resulting from searches for variation of the Newton's constant.
Subsequently, it will be confronted with the large-scale structure data.
We will show that up to scales of $\sim10^4\,$Mpc,
the derived bound on the spatial derivative of the matter energy density
does not allow for the existence of the large-scale structure as observed today. 

The rest of this paper is organized as follows: 
In section \ref{section_effEinstein} the derivation of the effective Einstein-like equation on the brane is discussed, 
whereas the detailed procedure can be found in Appendix A.\footnote{
  In Appendix B, a model-independent upper bound on the present value of the dilaton time derivative is derived.} 
The application of effective gravitational equations for the AdS$_5$ bulk,
with the matter content of the universe localized on the brane and described by a perfect fluid,
is presented in Section \ref{section_inhomogeneous}.
A constraint on the spatial derivative of the matter energy density is derived.
This result is quantified in section \ref{section_lateuniverse},
by employing observational limits from the time (non-)variation of the Newton's constant.
Subsequently, a comparison with observational data on large-scale structure 
is conducted in Section \ref{section_lssdata}.
Section \ref{section_conclusions} contains our conclusions.

\section{Effective Einstein-like equation at the brane}
\label{section_effEinstein}

As was discussed in the introduction,
a physically viable theory of higher order dilaton gravity - motivated by string theories - 
was introduced in Ref.\ \cite{paper2}.
Its equations of motion were constructed up to arbitrary order in derivatives 
of both the metric tensor and the dilaton field.
However, the search for novel features in the effective cosmological picture
should naturally start from considering the first order dilaton gravity (hereafter: dilaton gravity).
Its Lagrangian derived in Ref.\ \cite{paper2} in the string frame reads
\beq
 \widetilde{\cL} 
 = {\textstyle\frac{\alpha_1}{2}} \, e^{-\phi}  
    \left[ \widetilde{\cR} + 2 \, \widetilde{\nabla}^\sigma\widetilde{\pa}_\sigma\phi - (\widetilde{\pa}\phi)^2 \right]
 \, ,
 \label{Lagrangian_string}
\eeq
where $\alpha_1$ is a constant,
$\widetilde{\cR}$ and $\widetilde{\nabla}_\mu$ denote the Ricci scalar and the covariant derivative
associated with the 5-dimensional bulk metric $\widetilde{g}_{\mu\nu}$, respectively.
The first derivative $(\widetilde{\pa}_\mu\phi)$ was equipped with a tilde as well - 
in order to underline that the terms involving it will be also influenced by the conformal transformation
due to implicit contractions of indices.

Through a conformal rescaling, the Lagrangian (\ref{Lagrangian_string}) can be transformed to the Einstein frame.
By demanding that its gravitational part has to take the form of the Einstein-Hilbert Lagrangian in the Einstein frame, 
the conformal transformation relating these frames is found to be
\beq
 \widetilde{g}_{\mu\nu} 
  = e^{2\phi/3} \, g_{\mu\nu} \, .
\eeq
Thus, the main bulk quantities transform as
\bea
&&
 \widetilde{\cR} 
  = e^{-2\phi/3} \big[ \cR
    - {\textstyle\frac83} \, \nabla^\sigma\pa_\sigma\phi
    - {\textstyle\frac43} \, (\pa\phi)^2 \big]
 \, , 
\\[2pt]
&&
 \widetilde{\nabla}^\sigma\widetilde{\pa}_\sigma\phi
  = e^{-2\phi/3} \big[ \nabla^\sigma\pa_\sigma\phi
    + (\pa\phi)^2 \big]
 \, ,
\\[2pt]
&& 
 (\widetilde{\pa}\phi)^2
  = e^{-2\phi/3} \, (\pa\phi)^2
 \, ,
\eea
where $\cR$ and $\nabla_\mu$ denote the Ricci scalar and the covariant derivative
associated with the 5-dimensional bulk metric $g_{\mu\nu}$, respectively.
Hence, the 5-dimensional Lagrangian (\ref{Lagrangian_string}) of dilaton gravity 
in the Einstein frame reads
\beq
 \cL 
  = {\textstyle\frac{\alpha_1}{2}} \left[ 
    \cR
    - {\textstyle\frac23} \, \nabla^\sigma\pa_\sigma\phi
    - {\textstyle\frac13} \, (\pa\phi)^2 \right]
 \, .
 \label{Lagrangian_Einstein}
\eeq

As was already motivated in the introduction, 
it is interesting and important to consider what gravity would be induced at the brane
on which the matter content of the universe is localized
- i.e.\ the effective 4-dimensional description -
if the bulk action was given by a dilaton gravity theory (\ref{Lagrangian_Einstein}). 
The effective gravitational equations at the brane 
allow to study the effective cosmology in brane scenarios,
its dependence on the details of the bulk gravity theory,
and the modifications with respect to the standard picture 
arising from the 4-dimensional Einstein gravity.
Consequently, in the present work a 5-dimensional theory of dilaton gravity 
with brane-localized matter will be considered
- given by the following Lagrangian:
\beq
 \cL 
  = {\textstyle\frac{\alpha_1}{2}} \Big[ 
    \cR
    - {\textstyle\frac23} \nabla^\sigma\pa^{(5)}_\sigma\phi
    - {\textstyle\frac13} (\pa^{(5)}\phi)^2 \Big]
  - V(\phi)  
  + \big[ f(\phi) \cL_m + \lambda(\phi) \big] \delta_B
 \, , 
 \label{Lagrangian}
\eeq
where $V(\phi)$ is a scalar potential in the bulk,
$\cL_m$ stands for the matter content of the universe,
$f(\phi) \, \cL_m$ denotes the (non-minimal) coupling 
of the dilaton field $\phi$ to $\cL_m$,
$\lambda(\phi)$ can be identified as the `cosmological constant'-type term on the brane, 
and the Dirac delta type distribution $\delta_B$ yields the position of the 4-dimensional brane. 
The first part of the Lagrangian (\ref{Lagrangian}) 
is given by the theory (\ref{Lagrangian_Einstein}) of dilaton gravity.
The superscript in $\pa^{(5)}_\mu$ has been added 
to underline that this derivative is not projected on the brane,
whereas $\pa_\mu$ will denote the derivative projected on the brane.

In order to maintain the full generality, 
the effective equations of motion at the brane will be derived in the covariant approach. 
We will follow the procedure established in Ref.\ \cite{paper3}, 
the detailed calculations can be found in Appendix A.
Here, only the key points will be discussed 
in order to present the starting point for the next sections,
and clarify certain misconceptions in the literature. 

For the derivation of the effective gravitational equations at the brane in the covariant approach,
the crucial definition is that of the induced (projected) brane metric
\beq
 h_{\mu\nu}
  = g_{\mu\nu} - n_\mu n_\nu
 \, ,
 \label{h}
\eeq
where $n^\mu$ denotes a vector field (normalized to 1) perpendicular to the brane at its position.
It should be underlined here, that the metric tensors $g_{\mu\nu}$ and $h_{\mu\nu}$ 
have separate sets of basic quantities (Ricci scalar, covariant derivative) associated with them: 
$\cR$, $\nabla_\mu$,
and $R$, $D_\mu$, respectively.

The subsequent steps in the derivation of the effective gravitational equations at the brane include
calculation of the bulk equations of motion (\ref{tensor0})-(\ref{scalar0})
and of directional limits (i.e.\ evaluated `next' to the brane) (\ref{Thh})-(\ref{W})
of their projections on the brane and/or on the normal vector field $n^\mu$,
discussion of the junction condition (\ref{junction1})-(\ref{junction2}), 
as well as further calculations and definitions relevant for the procedure.
Thus, the detailed derivation of the effective equations of motion at the brane conducted in Appendix A
leads to the effective Einstein-like equation at the brane, reading
\begin{equation}
 R_{\mu\nu} 
  - {\textstyle\frac12} \, h_{\mu\nu} R 
  = 8\pi \bG(\phi) \, \tau_{\mu\nu}
  - h_{\mu\nu} \bL(\phi)
  + {\textstyle\frac{f^2(\phi)}{4\alpha_1^2}} \pi_{\mu\nu}
  - E_{\mu\nu}
  + {\textstyle\frac29} \, (\pa_\mu\phi) (\pa_\nu\phi)
  - {\textstyle\frac{5}{36}} \, h_{\mu\nu} \, (\pa\phi)^2 
 \, .
 \label{Eeq}
\end{equation}
The brane localized sources are defined as 
\beq 
 \tau_{\mu\nu} 
  = h_{\mu\nu} \, \cL_m - 2 \, {\textstyle\frac{\delta\cL_m}{\delta h^{\mu\nu}}}
  \, , \quad
 \tau_\phi 
  = {\textstyle\frac{f'(\phi)}{f(\phi)}} \, \cL_m + {\textstyle\frac{\delta\cL_m}{\delta\phi}}
 \, ,
 \label{taus}
\eeq
with the prime denoting the derivative with respect to the dilaton.
Terms quadratic in the brane energy-momentum tensor $\tau_{\mu\nu}$, which are typical of brane gravity, are collected in 
\beq
 \pi_{\mu\nu} 
  \equiv - \, \tau_{\mu\rho}\tau^\rho_\nu
      + {\textstyle\frac13} \, \tau \, \tau_{\mu\nu}
      + {\textstyle\frac12} \, h_{\mu\nu} \tau_\rho^\sigma\tau^\rho_\sigma
      - {\textstyle\frac16} \, h_{\mu\nu} \, \tau^2
 \, .
 \label{pi}
\eeq
The notation of effective Newton's and cosmological constants has been introduced,
namely
\bea
 \bG(\phi) 
  &=& {\textstyle\frac{-1}{48\pi\alpha_1^2}} f(\phi) \lambda(\phi)
 \, ,
 \label{G}
\\[2pt]
 \bL(\phi) 
  &=& {\textstyle\frac{1}{2\alpha_1}} V(\phi) - {\textstyle\frac{f(\phi)^2}{4\alpha_1^2}} \left[ 
      {\textstyle\frac{3}{4}} \tau_\phi^2 
      + {\textstyle\frac{3\lambda'(\phi)}{2f(\phi)}} \tau_\phi
      - {\textstyle\frac{\lambda(\phi)^2}{3f(\phi)^2}} 
      + {\textstyle\frac{3\lambda'(\phi)^2}{4f(\phi)^2}} \right]
 \, .
 \label{Lambda}
\eea
It should be underlined, that a single, but generically non-vanishing term\footnote{
  Note that there is no direct dependence on the bulk scalar solution.
  In Ref.\ \cite{MeBa} on the other hand, it is claimed that the dependence on $E_{\mu\nu}$
  of the effective brane equations is easier to be removed than that on the bulk scalar field.
  The detailed calculation in Appendix A proves otherwise.}
\beq
 E_{\mu\nu} 
  = n^\alpha h_\mu^\beta n^\gamma h_\nu^\delta \, \cC_{\alpha\beta\gamma\delta} 
 \, ,
 \label{E}
\eeq
where $\cC_{\alpha\beta\gamma\delta}$ stands for the bulk Weyl tensor,
denotes the explicit bulk contribution,
and thus yields the total bulk's influence on the brane gravity.
In order to fully describe the gravity induced on the brane,
the solution of the equations of motion for the bulk gravity would have to be known.

To summarize, the effective Einstein-like brane equation (\ref{Eeq})
involves three types of terms which can be treated as corrections to the standard Einstein equation:
terms quadratic in the brane localized sources (\ref{taus}) - typical of brane gravity theories,
kinetic terms for the dilaton - typical of scalar-tensor theories of gravity,
and - last but not least - the explicit bulk's contribution (\ref{E}).

Let us underline here the importance of assuming a $\Z2$ symmetry for the bulk
(with its fixed point at the brane's position). 
In the literature, this was always assumed, but never motivated.
However, at it was shown in Ref.\ \cite{paper3} and will be also pointed out in Appendix A,
this assumption is crucial for both the form and the existence of the effective gravitational equations at the brane.
In the absence of the bulk $\Z2$ symmetry, the following consistency condition (on the brane sources) 
would be the only effective brane equation obtained:
\beq
 D_\lambda \! \left( f(\phi) \, \tau_\mu^\lambda \right) 
  = f(\phi) \, \tau_\phi (\pa_\mu\phi)
 \, .
 \label{consistency}
\eeq
It can be also interpreted as a `generalized' covariant conservation of the energy-momentum tensor on the brane,
as in the case without the dilaton field, it reduces to the covariant conservation 
of the brane energy-momentum tensor $\tau_{\mu\nu}$.
It is the inclusion of the bulk $\Z2$ symmetry that guarantees the existence 
of a `true'\footnote{
  i.e.\ expressed exclusively in terms of brane quantities,
  relating the brane fields $h_{\mu\nu}$ and $\phi$ 
  to the brane localized sources $\tau_{\mu\nu}$ and $\tau_\phi$,
  and thus describing the dynamics of $h_{\mu\nu}$ and $\phi$} 
effective equation of motion at the brane (\ref{effeq}) 
- irrespective of the brane interactions' nature or other setup details.

Let us comment on the derivation of the effective brane equations in the Einstein frame (carried out in Appendix A)
as compared to the corresponding procedure in the string frame (established in Ref.\ \cite{paper3})
- from the point of view of the directional limit (\ref{W}) of the bulk equation of motion for the dilaton (\ref{scalar0}). 
The issue of the effective brane equation of motion for the scalar field 
caused some misunderstandings in the literature.
In the usually employed Einstein frame, 
it seems as if the information carried by this equation could not be extracted or employed.
However, as soon as the issue is addressed in the string frame (as in Ref.\ \cite{paper3}),
it turns out that this equation actually plays an important role 
in the derivation of the effective Einstein-like equation at the brane.
It seems that the apparent decoupling of the brane equation for the scalar field from the gravitational one
is a consequence of the specific character of the Einstein frame: the dilaton is minimally coupled to gravity,
whereas in the string frame the simultaneous treatment of both fields is essential.

\section{Search for inhomogeneities on the brane in an AdS$_5$ bulk}
\label{section_inhomogeneous}

In what follows, two commonly adopted assumptions will be taken.
In many models the spacetime is assumed to be of anti de Sitter type,
whose popularity is not only due to the AdS/CFT correspondence,
but also to its importance in higher-dimensional models (e.g.\ the Randall-Sundrum model),
as well as it being one of the simplest types of spacetimes.
Adopting an exact AdS$_5$ spacetime as the description of the bulk 
results in $E_{\mu\nu}=0$ according to the definition (\ref{E}),
and thus no bulk influence on the brane gravity according to the effective Einstein-like brane equation (\ref{Eeq}).
As for the matter content of the universe, 
in cosmological considerations the most typical description for the (dark) matter and radiation
is that of a perfect fluid, meaning that the matter energy-momentum tensor takes the form of
\beq
 \tau_{\mu\nu} 
  = \rho_m \, t_\mu t_\nu + p_m \, \gamma_{\mu\nu}
 \, ,
 \label{fluid}
\eeq
where the 3-dimensional spatial metric reads
$\gamma_{\mu\nu} = h_{\mu\nu} + t_\mu t_\nu$, \footnote{
  Note that $\gamma_{\mu\nu} t^\nu = 0$ and $t^\mu t_\mu = -1$.}
while $\rho_m$ and $p_m$ denote the energy density and the pressure of the (dark) matter and radiation on the brane, 
respectively.

Since the dilaton gravity brane scenario addressed here
can be regarded as an extended theory of gravity as motivated by string theories,
it is relevant to check whether for such a popular assumption as AdS$_5$ bulk,
any inhomogeneities in the perfect fluid - and thus the large-scale structure - are permitted on the brane.
The following ingredients will be employed to investigate this issue:
the effective gravitational (Einstein-like) equation at the brane (\ref{Eeq}) 
with $E_{\mu\nu}=0$ due to the AdS$_5$ bulk,
the consistency condition (\ref{consistency}) on the brane sources,
and the 4-dimensional Bianchi identity 
\beq
 D^\nu \big( R_{\mu\nu} - {\textstyle\frac12} h_{\mu\nu} R \big) = 0
 \, .
 \label{Bianchi}
\eeq
The perfect fluid (\ref{fluid}) will be substituted for the energy-momentum tensor $\tau_{\mu\nu}$ on the brane.

Consequently, a set of four equations at the brane is obtained, namely
\bea
&&\hspace{-1cm}
  \left[ \lambda \, \tau_\phi
    - \lambda' \rho_m
    + {\textstyle\frac{6\alpha_1^2}{f}} \overline\Lambda'
    + f' \rho_m^2 
    - {\textstyle\frac{4\alpha_1^2}{3f}} (D^\sigma\pa_\sigma\phi) \right] t^\nu (\pa_\nu\phi)
  + {\textstyle\frac{\alpha_1^2}{3f}} \, t^\rho (D^\nu\pa_\rho\phi) (\pa_\nu\phi) 
\nn[2pt]
&&\hspace{-1cm}
  \qquad + \, f \rho_m \left[ t_\nu \, \pa^\nu \! \rho_m 
    + \left( \rho_m + p_m \right) D^\nu t_\nu \right] 
 = 0  
 \, , 
 \label{set1}
 \\[2pt]
&&\hspace{-1cm}
  \left[ - \lambda \, \tau_\phi
    - \lambda' p_m 
    - {\textstyle\frac{6\alpha_1^2}{f}} \overline\Lambda'
    + f' \rho_m \left( \rho_m + 2 \, p_m \right)
    + {\textstyle\frac{4\alpha_1^2}{3f}} (D^\sigma\pa_\sigma\phi) \right] \gamma_\mu^\nu (\pa_\nu\phi)
  - {\textstyle\frac{\alpha_1^2}{3f}} \, \gamma_\mu^\rho (D^\nu\pa_\rho\phi) (\pa_\nu\phi) 
\nn[2pt]
&&\hspace{-1cm}
  \qquad + \, f \rho_m \left[ \gamma_{\mu\nu} \, \pa^\nu \! p_m + \left( \rho_m + p_m \right) t_\nu D^\nu t_\mu\right] 
  + f \left( \rho_m + p_m \right) \gamma_{\mu\nu} \, \pa^\nu \! \rho_m 
 = 0 
 \, , 
 \label{set2}
\\[2pt]
&&\hspace{-1cm}
 t^\nu \, \pa_\nu \rho_m
  + \left( \rho_m + p_m \right) D_\nu t^\nu
 = - \left( \tau_\phi + {\textstyle\frac{f'}{f}} \rho_m \right) t^\nu (\pa_\nu\phi) 
 \, ,
 \label{set3}
\\[2pt]
&&\hspace{-1cm}
  \gamma_{\mu\nu} \, \pa^\nu p_m
  + \left( \rho_m + p_m \right) t^\nu D_\nu t_\mu
 = \left( \tau_\phi - {\textstyle\frac{f'}{f}} p_m \right) \gamma_\mu^\nu (\pa_\nu\phi)
 \, .
 \label{set4}
\eea
To be more specific,
Eqs.\ (\ref{set1}) and (\ref{set2}) originate from the parts parallel and perpendicular to the vector field $t^\mu$
of the 4-dimensional Bianchi identity (\ref{Bianchi}) 
employed for the effective Einstein-like brane equation (\ref{Eeq})
with $E_{\mu\nu}=0$ and a perfect fluid on the brane (\ref{fluid}),
whereas Eqs.\ (\ref{set3}) and (\ref{set4}) are given by parts parallel and perpendicular to the vector field $t^\mu$
of the consistency condition at the brane (\ref{consistency}) evaluated for the perfect fluid (\ref{fluid}).

Combining Eq.\ (\ref{set1}) with Eq.\ (\ref{set3}), and Eq.\ (\ref{set2}) with Eq.\ (\ref{set4}), leads to:
\bea
&&\hspace{-2cm}
  \left[ - \left( \rho_m - {\textstyle\frac{\lambda}{f}} \right) \tau_\phi
    - {\textstyle\frac{\lambda'}{f}} \rho_m
    + {\textstyle\frac{6\alpha_1^2}{f^2}} \overline\Lambda'
    - {\textstyle\frac{4\alpha_1^2}{3f^2}} \, (D^\sigma\pa_\sigma\phi) \right] t^\nu (\pa_\nu\phi)
  + {\textstyle\frac{\alpha_1^2}{3f^2}} \, t^\rho (D^\nu\pa_\rho\phi) (\pa_\nu\phi)
 = 0  
 \, ,
\\[2pt]
&&\hspace{-2cm}
  \left[ \left( \rho_m - {\textstyle\frac{\lambda}{f}} \right) \tau_\phi
    - {\textstyle\frac{\lambda'}{f}} p_m 
    + {\textstyle\frac{f'}{f}} \rho_m \left( \rho_m + p_m \right)
    - {\textstyle\frac{6\alpha_1^2}{f^2}} \overline\Lambda'
    + {\textstyle\frac{4\alpha_1^2}{3f^2}} \, (D^\sigma\pa_\sigma\phi) \right] \gamma_\mu^\nu (\pa_\nu\phi)
\nn[2pt]
&&\hspace{-2cm}
  \qquad 
  - \, {\textstyle\frac{\alpha_1^2}{3f^2}} \, \gamma_\mu^\rho (D^\nu\pa_\rho\phi) (\pa_\nu\phi)
  + \left( \rho_m + p_m \right) \gamma_{\mu\nu} \, \pa^\nu \! \rho_m 
 = 0 \, .
\eea
Again, these two equations can be combined for $(t^\nu \pa_\nu\phi) \neq 0$.\footnote{
  $(t^\nu \pa_\nu\phi) = 0$ would imply a constant dilaton field. } 
Consequently, for the adopted assumptions of the AdS$_5$-type bulk 
and a perfect fluid description of the universe's matter content localized on the brane,
the spatial derivative of the energy density on the brane reads
\bea
&&\hspace{-1.25cm}
 \gamma_\mu^\nu \, \pa_\nu \rho_m 
  = - \left( {\textstyle\frac{f'}{f}} \rho_m - {\textstyle\frac{\lambda'}{f}} \right) \gamma_\mu^\nu (\pa_\nu\phi)
  + \, {\textstyle\frac{\alpha_1^2}{3f^2\left( \rho_m + p_m \right)}} \left[ 
    \gamma_\mu^\rho (D^\nu\pa_\rho\phi) (\pa_\nu\phi)
    - \frac{ t^\rho (D^\sigma\pa_\rho\phi) (\pa_\sigma\phi)}
      {t^\lambda (\pa_\lambda\phi)} \gamma_\mu^\nu (\pa_\nu\phi) \right]
 . \;
 \label{rhomifull}
\eea
This result imposes a strict condition on the matter content of the universe, 
localized on the 4-dimensional brane.
It relates any spatial inhomogeneities in the perfect fluid form of the energy-momentum tensor on the brane
- and thus in the (dark) matter and radiation - to the value of the spatial (brane) derivatives of the dilaton field.
Hence, Eq.\ (\ref{rhomifull}) can potentially put strong constraints 
on the allowed size of inhomogeneities in the matter distribution on the brane,
and thus on the cosmological large-scale structure as observed today.\footnote{
  It was shown in Ref.\ \cite{SaShMa} that for Einstein gravity in an exactly AdS$_5$ bulk,
  an inhomogeneous perfect fluid on the brane is rejected, 
  and thus only a perfectly spatially homogeneous universe would be allowed.}
In the following section, the observational limits on the (non-)variation of the Newton's constant
will be employed to quantify the implications of relation (\ref{rhomifull}).
The resulting constraint will be compared with the observational data on large-scale structure in Section \ref{section_lssdata}.

\section{Allowed inhomogeneities in the late universe}
\label{section_lateuniverse}

All results up to this point were obtained in the covariant approach, maintaining the full generality. 
As for the purpose of this section, it is more useful to simplify the notation.
The relation (\ref{rhomifull}) is first rewritten as
\beq
 \rho_m,_i
  = - \left( {\textstyle\frac{f'}{f}} \rho_m - {\textstyle\frac{\lambda'}{f}} \right) \phi,_i
  + \, {\textstyle\frac{\alpha_1^2}{3f^2\left( \rho_m + p_m \right)}} \left[ 
    D^\nu\pa_i\phi
    - \dot\phi^{-1} \phi,_i D^\nu\pa_t\phi \right] (\pa_\nu\phi) 
 \, ,
 \label{rhomi}
\eeq 
where the time and the spatial derivatives are denoted by an (over)dot and a comma, respectively,  
and the index $i$ stands for a spatial direction perpendicular to the vector field $n^\mu$,
i.e.\ a spatial direction in the hypersurface of the brane.
Moreover, from this point on it will be assumed that $\lambda\neq\lambda(\phi)$,
where $\lambda$ denotes the `cosmological constant'-type term\footnote{
  Note that $\lambda$ is only one of the contributions 
  to the effective brane cosmological constant $\overline\Lambda$ defined in Eq.\ (\ref{Lambda}).}
accompanying the brane matter term $\cL_m$ in the Lagrangian (\ref{Lagrangian}).
Consequently, $\lambda'=0$ will be put in Eq.\ (\ref{rhomi}).

The formula for the spatial derivative of the energy density (\ref{rhomi})
involves two types of terms - with either one, or three derivatives acting on the dilaton field. 
The latter can be neglected at least at late times, which can be seen as follows.
First, for the late universe the dilaton field is expected to be almost constant,
and a model-independent bound $\dot\phi_0 \lesssim 2.4 \, H_0 \simeq 1.8 \, \big( 10^{10}$ yr$\big)^{-1}$
set by current observational data is derived in Appendix B.\footnote{
  This bound bases on the deceleration parameter $q_0<0$,
  addressed for negligible $\rho_m^2$ corrections to the modified Friedmann equations.
  The assumptions include: fractional matter density $\Omega_{m0}>0.25$
  ($\Omega_i=\rho_i/\rho_c$, where $\rho_c$ is the critical density of a flat universe),
  spatially-flat universe with $\Omega_m + \Omega_{\overline\Lambda} +\Omega_\phi = 1$, 
  Hubble constant $H_0 \simeq 72$ km/(s\,Mpc) \cite{PDG},
  where the subscript 0 indicates the late universe values.
  For details see Appendix B.}
Second, it can be assumed that $|\ddot\phi_0|\ll\dot\phi_0^2$,  
as otherwise the currently observed $\phi_0 \approx {\rm const}$ would constitute a coincidence problem.
Third, for typical models involving a scalar field, we have\footnote{
 Any initial inhomogeneities of the dilaton would have been washed out by the inflation.} 
\beq
 |\phi,_i| 
  \lesssim c_1 |\dot\phi|
 \, ,
 \label{c1}
\eeq 
where $c_1$ is positive and of order 1. 
Consequently, in Eq.\ (\ref{rhomi}) terms ${\cal O} \big( (\pa\phi) \big)$ will be treated as still non-negligible,
whereas terms ${\cal O} \big( (\pa\phi) (D\pa\phi) \big)$ will be dropped. 

Hence, for the late universe, Eq.\ (\ref{rhomi}) reduces to a very simple form of
\beq
 \rho_{m0},_i
  \simeq - {\textstyle\frac{f'}{f}} \rho_{m0} \, \phi_0,_i 
 \, ,
 \label{rhomi0}
\eeq 
which means that the spatial inhomogeneities in the matter energy density $\rho_m$ on the brane are highly constrained
for the popular assumptions of an AdS$_5$-type bulk 
and the brane matter energy-momentum tensor of the perfect fluid form (\ref{fluid}).
In particular, an inhomogeneous perfect fluid (i.e.\ with $\rho_m,_i \neq 0$) on the brane 
is allowed only if the matter on the brane is non-minimally coupled to the dilaton (so that $f'\neq0$).
Moreover, a perfectly spatially homogeneous dilaton field would clearly imply the lack of any matter inhomogeneities.

In order to quantify the constraint (\ref{rhomi0}) 
on the spatial derivative of the brane matter energy density $\rho_{m0}$ in the late universe,
current observational limits from the time (non-)variation of the Newton's constant can be employed
- in combination with the constraint (\ref{c1}) 
on the spatial derivative of the dilaton as compared to its time derivative.
The most recent bounds on the relative time variation $|\dot{G}/G|$ of the Newton's constant $G$
are summarized in Ref.\ \cite{Uz}.
They arise from solar system constraints, pulsar timing, stellar constraints, 
and cosmological constraints.
Most of those results reach as far as $|\dot{G}_0/G_0|<\big( 10^{11}\,\textrm{yr} \big)^{-1}$,
thus allowing to put a stringent constraint
on the time variation of the effective Newton's constant (\ref{G}) reading
\beq
 \big| \dot{\bG}_0 / \bG_0 \big|
  < \big( 10^{11}\,\textrm{yr} \big)^{-1}
 \, .
 \label{dotGobs}
\eeq
Recalling the formula for the effective brane Newton's constant (\ref{G}), 
it directly follows that 
\beq
 \big| \pa_\mu \bG_0 / \bG_0 \big|
    = \big| \textstyle{\frac{f'}{f}} \big| \, (\pa_\mu\phi_0)
\eeq
for $\lambda \neq \lambda(\phi)$.
Hence the observational limit (\ref{dotGobs}) can be rewritten as
$\big| \frac{f'}{f} \dot\phi_0 \big| < \big( 10^{11} \, \textrm{yr} \big)^{-1}$.
Employing the constraint (\ref{c1}) on the spatial derivative of the dilaton with respect to its time derivative, 
a limit of $\big| \frac{f'}{f} \phi_0,_i \big| \lesssim 3.3 \, c_1 \big( 10^{5} \, \textrm{Mpc} \big)^{-1}$ is obtained.

Consequently, using the relation (\ref{rhomi0}) a following constraint is obtained
for the spatial derivative of the brane matter energy density:  
\beq
 |\rho_{m0},_i|
  \lesssim 3.3 \, c_1 \, \rho_{m0} \big( 10^5 \, \textrm{Mpc} \big)^{-1}
 \, ,
 \label{rhomi0limit}
\eeq 
whose full implications will be demonstrated in the following section  
through a comparison with the observational data on large-scale structure.

\section{Late universe: inhomogeneities and large-scale structure}
\label{section_lssdata}

Detailed studies of the large-scale structure of the universe,
i.e.\ the spatial distribution of galaxies, their groups and clusters,
are capable of providing information on such model-dependent issues 
as the overall matter distribution\footnote{
  Although in principle only the visible part of baryonic matter is observed, 
  it is usually assumed that the overall baryonic matter distribution is probed in experiments. 
} or the physics of galaxy formation.
The distribution of galaxies in space is not random, but shows a variety of structures.
The content and the statistical properties of the large-scale structure are addressed
by large galaxy redshift surveys, which probe the galaxy distribution.
These include such experiments as e.g.\ the Sloan Digital Sky Survey (SDSS) \cite{SDSS} 
or the two-degree Field Galaxy Redshift Survey (2dFGRS) \cite{2dFGRS}.
The spatial distribution of galaxies is typically characterized statistically
through the so-called two-point correlation function $\xi(x)$,
which can be interpreted as the excess number of galaxy pairs of a given separation $x$, 
relative to that expected for a random distribution.

Following the standard approach presented e.g.\ in Ref.\ \cite{book},
a statistical description of cosmological perturbations can be established
in order to relate theory to observations.
The statistical properties of a random density perturbation field (density contrast)
\beq
 \delta(\vec{x},t)
  \equiv \frac{\rho(\vec{x},t)}{\rho_{av}(t)} - 1
 \, ,
 \label{delta}
\eeq
where $\rho_{av}(t)$ denotes the mean density,
are characterized by the probability distribution function
${\cal P}_x(\delta_1,\delta_2,\ldots,\delta_n)\ud\delta_1\ud\delta_2\ldots\ud\delta_n$,
which gives the probability that the $\delta$ field values at positions $\vec{x_i}$
are in the range $[\delta_i,\delta_i+\ud\delta_i]$, with $i=1,2,\ldots,n$.
The function ${\cal P}_x$ is fully characterized by its moments
\beq
 \langle\delta_1^{l_1}\delta_2^{l_2}\cdots\delta_n^{l_n}\rangle
  = \int \delta_1^{l_1}\delta_2^{l_2}\cdots\delta_n^{l_n} 
    {\cal P}_x(\delta_1,\delta_2,\ldots,\delta_n)\ud\delta_1\ud\delta_2\ldots\ud\delta_n
 \, ,
\eeq
where $l_i \in {\mathbb Z}$ and $l_i\geq 0$.
In particular, $\langle\delta(\vec{x})\rangle = 0$, and
\beq
 \xi(x) 
  = \langle \delta_1 \delta_2 \rangle
  \label{2point}
\eeq
is the two-point correlation function with $x \equiv | \vec{x}_1 - \vec{x}_2 |$.
Hence, the variance $\sigma^2 = \langle \delta^2(\vec{x}) \rangle = \xi(0)$
of the perturbation field does not depend on $\vec{x}$. 

In this section, the constraint (\ref{rhomi0limit}) 
on the allowed values of the spatial derivative of the brane matter energy density,
predicted by the studied dilaton gravity scenario in AdS$_5$ bulk, 
will be compared with the observational data on the large-scale structure of the universe.
Since it is relatively straightforward to measure,
the two-point correlation function (\ref{2point}) of galaxies has been estimated 
from various samples within galaxy surveys.\footnote{
  However, the data is usually presented in the form of the power spectrum of the perturbation field 
  - i.e.\ the Fourier transformation of the two-point correlation function (\ref{2point}).}
As the aim of the following calculation is just an estimation, 
allowing to compare the model's prediction (\ref{rhomi0limit}) with the observational data,
let me approximate the spatial derivative of the baryonic matter density 
by a ratio of the difference between the matter densities at two points
and the distance between these points. 
Consequently, the spatial derivative of matter energy density can be estimated 
from the large-scale structure as
\beq
 \langle |\rho,_x|^2 \rangle
  \simeq 
  \frac{2\rho_{av}^2\big( \xi(0) - \xi(x) \big)}{x^2}
 \, ,
 \label{lss}
\eeq
where as the variance $\xi(0)=\sigma^2$ 
usually $\sigma_8^2 \equiv \langle \delta_R^2(x) \rangle$ is taken,
i.e.\ the expected root mean square overdensity in a sphere of radius $R=8h^{-1}$Mpc $\simeq11.1\,$Mpc.
Here $\sigma_8=0.80(4)$ \cite{PDG} 
will be adopted.\footnote{
  Although the corresponding analysis of Wilkinson Microwave Anisotropy Probe (WMAP) 5-year results 
  involved the assumption of a flat $\Lambda$CDM universe
  with a power law initial spectrum, this value of $\sigma_8$ can be treated here as a reasonably good approximation.
  Note: the same applies to the values of $\Omega_m$ and $\Omega_b$.}

The relation (\ref{lss}) allows to estimate the value of the spatial derivative squared 
of the energy density of baryonic matter,
whereas the model's prediction (\ref{rhomi0limit}) applies for all matter in the universe.
Regarding that for most models of dark matter, its distribution is expected 
to be similar to that of baryonic matter, 
in the following calculation $\rho_m\simeq6\,\rho$ will be assumed.\footnote{
  Current observational values of the respective fractional densities read 
  $\Omega_m=0.26(2)$ and \mbox{$\Omega = \Omega_b = 0.044(4)$ \cite{PDG}.}}
With such a substitution, the limit (\ref{rhomi0limit}) on the spatial derivative of matter density 
predicted by the studied dilaton gravity scenario yields
\beq
 \langle |\rho,_i|^2 \rangle
  \lesssim \frac{0.1 \, c_1^2 \, \rho_{av}^2 \, (1+\xi(0))}{(10^4 \textrm{Mpc})^2} 
 \, ,
 \label{rhomi0lss}
\eeq
where $\langle |\rho,_i|^2 \rangle$ should be similarly understood 
as the `smearing' of $\langle \delta^2(\vec{x}) \rangle$ into $\sigma_8^2$.
  
Substituting $\sigma_8^2\simeq0.64$ for $\xi(0)$, 
the limit (\ref{rhomi0lss}) predicted by the dilaton gravity scenario can be now confronted 
with the estimate (\ref{lss}) based on the observational data.
Let us recall the main features of the two-point correlation function (\ref{2point}) 
as measured by e.g.\ the SDSS experiment \cite{plot}.
Usually $\xi(x)$ is plotted down to $x=10-60\,$Mpc only,
with its value estimated at a maximum of $\sigma_8^2$ when approaching $x=0$.
According to Ref.\ \cite{plot}, where $\xi(x)$ is plotted down to $x\simeq55\,$Mpc,
we have $\xi(55\,\textrm{Mpc})\simeq0.07$,
hence within the entire range of measured scales,
the numerator in Eq.\ (\ref{rhomi0lss}) is roughly an order of magnitude smaller 
than the numerator in Eq.\ (\ref{lss}). 
Furthermore, over a large range of scales (up to $x\sim10^4\,$Mpc basically),
the denominators differ strongly - by up to even 4 orders of magnitude for $x\simeq55\,$Mpc.
Thus, up to the scales where the two-point correlation function is consistent with 0,
the limit (\ref{rhomi0lss}) predicted by the model is much smaller 
than the estimate (\ref{lss}) based on the observational data.
Consequently, the today's large-scale structure cannot be described 
within the brane scenario of dilaton gravity (\ref{Lagrangian})
with an AdS$_5$ bulk and the perfect fluid description of the matter content of the universe.

\section{ Conclusions }
\label{section_conclusions}

In this work a theory of dilaton gravity (as motivated by string theories)
was addressed in a 5-dimensional brane scenario in the Einstein frame,
with a non-minimal coupling $f(\phi)\cL_m$ of the dilaton field to the matter content of the universe localized on the brane.
We introduced the non-minimal coupling to take into account the lack of consensus in the literature as to which of the conformally related frames
should be considered as the true physical framework for such a theory 
- and in which frame the dilaton-matter coupling should be minimal.

We started with deriving the effective gravitational equations at the brane in the covariant approach.
Contrary to the usual expectations, the analysis in the Einstein frame 
was technically more straightforward than in the string frame adopted in Ref.\ \cite{paper3}.
We addressed certain misconceptions in the literature, 
e.g.\ the interpretation of the information carried by the directional limit 
(i.e.\ evaluated `next' to the brane) of the bulk scalar equation,
the importance of $\Z2$ symmetry in the bulk (with its fixed point at the brane's position),
the non-vanishing bulk's influence on the brane gravity with no direct dependence on the bulk scalar solution.

In order to investigate basic properties of this dilaton gravity brane scenario,
we studied the implications of two commonly adopted assumptions: 
the bulk spacetime of an anti de Sitter type,
and a perfect fluid description for the matter content of the universe localized on the brane.
Employing only the effective Einstein-like brane equation,
the consistency condition on the brane sources and the 4-dimensional Bianchi identity,
we derived a strict constraint on the allowed values of the spatial derivative of the brane matter energy density.
We showed that apart from the earliest epochs of the universe's evolution,
the inhomogeneities in the brane matter are allowed 
only if the matter Lagrangian is non-minimally coupled to the dilaton in the Einstein frame, 
which is by no means a typical assumption. 
Furthermore, the inhomogeneities are proportional to the spatial derivative of the dilaton, and thus considerably suppressed 
- according to the model-independent upper bound (based on the current observational data) 
we obtained for the dilaton's time derivative. 

Employing the present limits from the time (non-)variation of the Newton's constant,
we rewrote our theoretical constraint on the spatial derivative of the matter density into an upper bound.
Subsequently, we confronted it with the observational data on large-scale structure from galaxy surveys.
We showed that up to scales of $\sim10^4\,$Mpc, the large-scale structure as we observe it today
is not allowed in this dilaton gravity brane scenario - at least for an exactly AdS$_5$ bulk
and the matter content of the universe described by perfect fluid.
This result holds not only for anti de Sitter type bulk,
but for any highly-symmetrical spacetimes with vanishing Weyl tensor.
Allowing for bulk deviations from an exact anti de Sitter spacetime (or other highly-symmetrical spacetime)
might enable a description of our universe in this string theory motivated scenario, 
as then the constraint on the spatial derivative of the brane matter would involve also the bulk Weyl tensor.
The situation might be also relaxed by including (into the dilaton gravity Lagrangian) 
terms of the higher order in derivatives of the metric tensor and the dilaton field - as motivated by string theories.
Verifying which of the specific models existing in the literature are affected
by the results of the present work lies beyond the scope of this paper.

\section*{Acknowledgments}

D.K.\ would like to thank Thomas Flacke, Thorsten Ohl, Marek Olechowski, Dominik Schwarz and Hermano Velten
for inspiring discussions and useful comments.

\section*{Appendix A}
\label{derivation}
\renewcommand{\theequation}{A.\arabic{equation}}
\setcounter{equation}{0}

As for the main steps of the derivation of effective equations of motion at the brane,
the procedure presented here follows the one introduced in Ref.\ \cite{paper3}.
Important differences arising between these two calculations 
due to the use of different conformal frames will be pointed out and discussed. 

In the theory of dilaton gravity given by the Lagrangian (\ref{Lagrangian}),
the (5-dimensional) bulk tensor $T_{\mu\nu}=0$ and scalar $W=0$ equations of motion 
read\footnote{
  Let us recall the notation: 
  the Ricci tensor $\cR_{\mu\nu}$ ($R_{\mu\nu}$)
  and the covariant derivative $\nabla_\mu$ ($D_\mu$) 
  are associated with the metric tensor $g_{\mu\nu}$ ($h_{\mu\nu}$ restricted to the brane's position),
  whereas $\pa_\mu$ ($\pa^{(5)}_\mu$) denotes the derivative (not) projected on the brane.}
\bea
&&\hspace{-0.5cm}
 \cR_{\mu\nu}
  - {\textstyle\frac12} \, g_{\mu\nu} \cR
  - {\textstyle\frac13} \, (\pa_\mu^{(5)}\phi)(\pa_\nu^{(5)}\phi)
  + {\textstyle\frac16} \, g_{\mu\nu} (\pa^{(5)}\phi)^2 
  + g_{\mu\nu} {\textstyle\frac{1}{\alpha_1}} V(\phi)  
\nn[2pt]
&&\hspace{2.05cm}
  - \, {\textstyle\frac{1}{\alpha_1}} \big[ f(\phi) \tau_{\mu\nu}
    + \lambda(\phi) h_{\mu\nu} \big] \delta_B
  = 0
 \, ,
 \label{tensor0}
\\[2pt]
&&\hspace{-0.5cm}
\nabla^\sigma\pa_\sigma^{(5)}\phi
  - {\textstyle\frac{3}{\alpha_1}} V'(\phi)  
  + {\textstyle\frac{3}{\alpha_1}} \big[ f(\phi) \tau_\phi
    + \lambda'(\phi) \big] \delta_B
  = 0
 \, ,
 \label{scalar0}
\eea
where $T_{\mu\nu} = \frac{1}{\sqrt{-g}} \frac{\delta}{\delta g^{\mu\nu}} \left( \sqrt{-g} \, \cL \right)$ 
and $W = \frac{1}{\sqrt{-g}} \frac{\delta}{\delta\phi} \left( \sqrt{-g} \, \cL \right)$.
The brane localized sources $\tau_{\mu\nu}$ and $\tau_\phi$ are defined in Eq.\ (\ref{taus}).

In order to derive the effective 4-dimensional equations of motion at the brane,
the quantities contributing to the bulk equations of motion (\ref{tensor0})-(\ref{scalar0})
have to be separated into parts parallel and perpendicular
to the vector field $n^\mu$.\footnote{
  As introduced in Section \ref{section_effEinstein}, the (normalized to 1) vector field $n^\mu$
  is perpendicular to the brane at its position.}
With the induced brane metric tensor defined in Eq.\ (\ref{h}),\footnote{
  Note that this expression holds at any point in the bulk,
  and yields the metric induced on the brane when restricted to the position of the brane.} 
the corresponding expressions are derived as
\bea
 \cR_{\mu\nu}  
  &=& R_{\mu\nu} - K K_{\mu\nu} + K_\mu^\sigma K_{\sigma\nu} 
  - \big( \bLn K_{\mu\nu} - K_\mu^\sigma K_{\sigma\nu} \big) 
\nn[2pt]
&&
  - \, n_\mu D_\nu K - n_\nu D_\mu K
  + n_\mu D_\lambda K_\nu^\lambda + n_\nu D_\lambda K^\lambda_\mu
  - n_\mu n_\nu \left( h^\sigma_\tau \bLn K_\sigma^\tau - K^\sigma_\tau K_\sigma^\tau \right)
 \, ,
 \label{decomposition1}
\\[2pt]
 \nabla_\mu\pa^{(5)}_\nu\phi 
  &=& D_\mu \pa_\nu\phi + K_{\mu\nu} \bLn\phi 
  + n_\mu n_\nu \big( \bLn^2 \phi - a^\lambda \pa^{(5)}_\lambda\phi \big)
\nn[2pt]
&&
  + \, n_\mu D_\nu \bLn\phi + n_\nu D_\mu \bLn\phi 
  - n_\mu K_\nu^\lambda (\pa_\lambda\phi) - n_\nu K_\mu^\lambda (\pa_\lambda\phi)
 \, , 
 \label{decomposition2}
\\[2pt]
 \pa^{(5)}_\mu\phi 
  &=& \pa_\mu\phi + n_\mu \bLn\phi
 \, ,
 \label{decomposition3}
\eea
where $\Ln$ is the Lie derivative\footnote{
  For a given tensor $M^{\rho_1\rho_2\cdots\rho_m}_{\sigma_1\sigma_2\cdots\sigma_l}$ 
  and arbitrary direction $v^\mu$, 
  the Lie derivative along $v^\mu$ is defined as follows:
  $\pounds_v  \, M^{\rho_1\rho_2\cdots\rho_m}_{\sigma_1\sigma_2\cdots\sigma_l}
    = v^\lambda \nabla_\lambda M^{\rho_1\rho_2\cdots\rho_m}_{\sigma_1\sigma_2\cdots\sigma_l}
    - \sum_{i=1}^m M^{\rho_1\rho_2\cdots\lambda\cdots\rho_m} _{\sigma_1\sigma_2\cdot\cdot\cdots\cdots\sigma_l}
      \nabla_\lambda v^{\rho_i}
    + \sum_{j=1}^l M^{\rho_1\rho_2\cdots\cdot\cdot\cdots\rho_m}_{\sigma_1\sigma_2\cdots\lambda\cdots\sigma_l}
      \nabla_{\sigma_j} v^\lambda$, 
  thus e.g.\ $\Ln\phi=n^\lambda \pa^{(5)}_\lambda \phi$.} 
along $n^\mu$, 
$K_{\mu\nu}=\frac12\Ln h_{\mu\nu}$ is the extrinsic curvature of hypersurfaces orthogonal to $n^\mu$,
and $a^{\lambda}\equiv n^\rho \nabla_\rho n^\lambda$.\footnote{Note that apart from the term 
  $a^\lambda\pa^{(5)}_\lambda\phi=n^\rho(\nabla_\rho n^\lambda)(\pa^{(5)}_\lambda\phi)$,
  the right-hand sides of the decompositions (\ref{decomposition1})-(\ref{decomposition3})
  involve only either brane quantities (which are orthogonal to $n^\nu$),
  or Lie derivatives along $n^\mu$ and $n^\mu$ itself.
  However, the term $a^\lambda\pa^{(5)}_\lambda\phi$ will not appear in the final results.}
Employing the dimensional reduction formulae (\ref{decomposition1})-(\ref{decomposition3})
and the definition (\ref{h}) of the induced brane metric tensor $h_{\mu\nu}$,
we can rewrite the bulk equations of motion (\ref{tensor0})-(\ref{scalar0}) as follows:
\bea
&&\hspace{-1.5cm}
 R_{\mu\nu}
  - {\textstyle\frac12} \, h_{\mu\nu} R
  + K_\mu K_\nu
  - K K_{\mu\nu}
  - {\textstyle\frac12} \, h_{\mu\nu} K^\sigma_\tau K_\sigma^\tau 
  + {\textstyle\frac12} \, h_{\mu\nu} K^2
\nn[2pt]
&&\hspace{-0.5cm}
  - \, \big( \bLn K_{\mu\nu} - K_\mu^\sigma K_{\sigma\nu} \big)
  + h_{\mu\nu} \big( h^\sigma_\tau \bLn K_\sigma^\tau - K^\sigma_\tau K_\sigma^\tau \big)
\nn[2pt]
&&\hspace{-0.5cm}
  - \, {\textstyle\frac13} \, (\pa_\mu\phi) (\pa_\nu\phi)
  + {\textstyle\frac16} \, h_{\mu\nu} (\pa\phi)^2
  + {\textstyle\frac16} \, h_{\mu\nu} (\bLn\phi)^2
  + h_{\mu\nu} {\textstyle\frac{1}{\alpha_1}} V(\phi) 
  - {\textstyle\frac{1}{\alpha_1}} f(\phi) \, \left( \tau_{\mu\nu} 
    + {\textstyle\frac{\lambda(\phi)}{f(\phi)}} h_{\mu\nu} \right) \delta_B
\nn[2pt]
&&\hspace{-0.5cm}
  + \, n_\mu \Big( 
    D_\lambda K_\nu^\lambda
    - D_\nu K
    - {\textstyle\frac13} \, (\bLn\phi) (\pa_\nu\phi) \Big)
  + n_\nu \Big( 
    D_\lambda K^\lambda_\mu
    - D_\mu K 
    - {\textstyle\frac13} \, (\bLn\phi) (\pa_\mu\phi) \Big)
\nn[2pt]
&&\hspace{-0.5cm}
  + \, n_\mu n_\nu \Big( 
    - {\textstyle\frac12} \big( R
      + K^\sigma_\tau K_\sigma^\tau 
      - K^2 \big)
    - {\textstyle\frac16} \, (\bLn\phi)^2
    + {\textstyle\frac16} \, (\pa\phi)^2
    + {\textstyle\frac{1}{\alpha_1}} V(\phi) \Big)
 = 0
 \, ,
 \label{tensor1}
\\
&&\hspace{-1.5cm}
 D^\sigma\pa_\sigma\phi 
  + K \bLn\phi 
  + \left( \bLn^2 \phi - a^\lambda \pa^{(5)}_\lambda\phi \right)
  - {\textstyle\frac{3}{\alpha_1}} V'(\phi)  
  + {\textstyle\frac{3}{\alpha_1}} f(\phi) \, \left( \tau_\phi 
    + {\textstyle\frac{\lambda'(\phi)}{f(\phi)}} \right) \delta_B
 = 0 
 \, .
 \label{scalar1}
\eea

Apart from brane related quantities, 
Eqs.\ (\ref{tensor1})-(\ref{scalar1}) involve terms discontinuous 
($K_{\mu\nu}$, $\bLn\phi$)
or singular ($\bLn K_{\mu\nu}$, $\bLn^2\phi$) at the brane's position.
In order to address them appropriately and take into account the information from all these contributions,
we should first consider what junction conditions have to be fulfilled at the brane.
Following the approach formulated in Ref.\ \cite{paper3}, where the necessary definitions can be found,
the junction conditions are obtained here by a 1-dimensional, infinitesimal, across-the-brane integration 
of the bulk equations of motion (\ref{tensor1})-(\ref{scalar1}) in the direction perpendicular to the brane:
\bea
 \big[ K_{\mu\nu} \big]_\pm
  &=& {\textstyle\frac{-1}{\alpha_1}} f(\phi) \left( \tau_{\mu\nu} 
    - {\textstyle\frac13} \, h_{\mu\nu} \, \tau 
    - {\textstyle\frac13} \, {\textstyle\frac{\lambda(\phi)}{f(\phi)}} h_{\mu\nu} \right)
 \, ,
 \label{junction1}
\\[2pt]
 \big[ \bLn\phi \big]_\pm
  &=& {\textstyle\frac{-3}{\alpha_1}} f(\phi) \, \left( \tau_\phi 
    + {\textstyle\frac{\lambda'(\phi)}{f(\phi)}} \right)
 \, , 
 \label{junction2}
\eea
where $[\cdot]_\pm = [\cdot]_+ - [\cdot]_-$,
with $[\cdot]_{+/-}$ denoting the limits of a given bulk quantity 
when approaching the brane from the `+' and the `-' sides.
The junction conditions (\ref{junction1})-(\ref{junction2}) 
determine across-the brane jumps in the values of the extrinsic curvature $K_{\mu\nu}$
and the Lie derivative of the dilaton field $\bLn\phi$,
which are caused by the presence of brane localized terms $\tau_{\mu\nu}$ and $\tau_\phi$.\footnote{
  Note that contrary to the junction conditions in the string frame, as obtained in Ref.\ \cite{paper3},
  the tensor junction condition (\ref{junction1}) involves only $\tau_{\mu\nu}$,
  and the scalar junction condition (\ref{junction2}) depends only on $\tau_\phi$.
  It is only due to the non-minimal dilaton-$\cL_m$ coupling function $f(\phi)$, 
  that the junction conditions related to $h_{\mu\nu}$ and $\phi$ are not fully separate,
  as expected for the Einstein frame.}

The effective brane equations of motion should follow from the bulk equations of motion,
and describe the dynamics of brane quantities - either defined exactly on the brane, 
or infinitesimally close to it.\footnote{
  For the detailed discussion of how the effective brane equations should be defined, see Ref.\ \cite{paper3}.}
Hence, let us write down the directional limits 
(i.e.\ when approaching the brane from the `+' or the `-' side)
of the bulk equations (\ref{tensor1})-(\ref{scalar1}).
We obtain the following equations:
\bea
&&\hspace{-2cm}
 \Big[ R_{\rho\sigma}
  - {\textstyle\frac12} \, h_{\rho\sigma} R
  + K_\rho K_\sigma
  - K K_{\rho\sigma}
  - {\textstyle\frac12} \, h_{\rho\sigma} K^\kappa_\lambda K_\kappa^\lambda 
  + {\textstyle\frac12} \, h_{\rho\sigma} K^2
\nn[2pt]
&&\hspace{-1.25cm}
  - \, \big( \bLn K_{\rho\sigma} - K_\rho^\tau K_{\tau\sigma} \big)
  + h_{\rho\sigma} \big( h^\kappa_\lambda \bLn K_\kappa^\lambda - K^\kappa_\lambda K_\kappa^\lambda \big) 
\nn[2pt]
&&\hspace{-1.25cm}
  - \, {\textstyle\frac13} \, (\pa_\rho\phi) (\pa_\sigma\phi)
  + {\textstyle\frac16} \, h_{\rho\sigma} (\pa\phi)^2
  + {\textstyle\frac16} \, h_{\rho\sigma} (\bLn\phi)^2
  + h_{\rho\sigma} {\textstyle\frac{1}{\alpha_1}} V(\phi) \Big]_{+/-}
 = 0
 \, ,
 \label{Thh}
\\[2pt]
&&\hspace{-2cm}
 \Big[ D_\lambda K^\lambda_\rho
  - D_\rho K 
  - {\textstyle\frac13} \, (\bLn\phi) (\pa_\rho\phi) \Big]_{+/-}
 = 0
 \, ,
 \label{Thn}
\\[2pt]
&&\hspace{-2cm}
 \Big[ \big( R
    + K^\sigma_\tau K_\sigma^\tau 
    - K^2 \big)
  + {\textstyle\frac13} \, (\bLn\phi)^2
  - {\textstyle\frac13} \, (\pa\phi)^2
  - {\textstyle\frac{2}{\alpha_1}} V(\phi) \Big]_{+/-}
 = 0
 \, ,
 \label{Tnn}
\\
&&\hspace{-2cm}
 \Big[ D^\sigma\pa_\sigma\phi 
  + K \bLn\phi 
  + \left( \bLn^2 \phi - a^\lambda \pa^{(5)}_\lambda\phi \right)
  - {\textstyle\frac{3}{\alpha_1}} V'(\phi) \Big]_{+/-}
 = 0 
 \, ,
 \label{W}
\eea
where Eqs.\ (\ref{Thh}), (\ref{Thn}), (\ref{Tnn})
were obtained by projecting the bulk tensor equation of motion (\ref{tensor1})
on the brane's hypersurface and/or on the normal vector field $n^\mu$
(i.e.\ by multiplying Eq.\ (\ref{tensor1}) by $h_\rho^\mu h_\sigma^\nu$, $h_\rho^\mu n^\nu$ and $n^\mu n^\nu$, respectively),
and Eq.\ (\ref{W}) arises from the bulk scalar equations of motion (\ref{scalar1}).

Eq.\ (\ref{Thh}) - i.e.\ the directional limit of the bulk tensor equation projected on the brane -
would be a natural candidate for an effective gravitational equation at the brane.
However, it involves $\bLn K_{\mu\nu}$, i.e.\ the second Lie derivative on the induced brane metric $h_{\mu\nu}$ 
in the direction perpendicular to the brane.
Let us then rewrite it, so that the explicit dependence on $\bLn K_{\mu\nu}$ is removed.
From the trace of Eq.\ (\ref{Thh}), the trace of the combination $\big( \bLn K_{\rho\sigma} - K_\rho^\tau K_{\tau\sigma} \big)$
can be determined, reading
\beq
 \big( h^\kappa_\lambda \bLn K_\kappa^\lambda - K^\kappa_\lambda K_\kappa^\lambda \big)
  = {\textstyle\frac13} \left[ R
    + K^\kappa_\lambda K_\kappa^\lambda
    - K^2
    - {\textstyle\frac23} \, (\bLn\phi)^2  
    - {\textstyle\frac13} \, (\pa\phi)^2
    - {\textstyle\frac{4}{\alpha_1}} V(\phi) \right]
 \, .
 \label{hLnK}
\eeq
On the other hand, by combining the decomposition (\ref{decomposition1}) 
with the definition of the 5-dimensional (bulk) Weyl tensor	
\beq
 \cC_{\mu\nu\rho\sigma}
  = \cR_{\mu\nu\rho\sigma}
  - {\textstyle\frac23} \big( 
    g_{\mu[\rho} \cR_{\sigma]\nu}
    - g_{\nu[\rho} \cR_{\sigma]\mu} \big)
  + {\textstyle\frac16} \, g_{\nu[\rho} g_{\sigma]\nu} \cR
 \, ,
 \label{Weyl}
\eeq
we can express the Lie derivative of the extrinsic curvature, $\bLn K_{\mu\nu}$,
as a function of its trace $h^\kappa_\lambda \bLn K_\kappa^\lambda$, namely
\bea
 \bLn K_{\mu\nu} - K_\mu^\sigma K_{\sigma\nu} 
    &=& {\textstyle\frac14} \, h_{\mu\nu} \big( h^\kappa_\lambda \bLn K_\kappa^\lambda - K^\kappa_\lambda K_\kappa^\lambda \big)
    - {\textstyle\frac12} \big( R_{\mu\nu} - K K_{\mu\nu} + K_\mu^\sigma K_{\sigma\nu} \big)
\nn[2pt]
&&
    + \, {\textstyle\frac18} \, h_{\mu\nu} \big( R - K^2 + K_\tau^\sigma K_\sigma^\tau \big)
    - {\textstyle\frac32} \, E_{\mu\nu}
 \, ,
 \label{LnK}
\eea
where $E_{\mu\nu}$ denotes the bulk Weyl tensor (\ref{Weyl}) projected on the brane, as defined in Eq.\ (\ref{E}).
Thus, employing Eqs.\ (\ref{LnK}) and (\ref{hLnK}), 
the tensor equation (\ref{Thh}) can be rewritten as follows:\footnote{
  Note that although the directional limit (\ref{W}) of the bulk scalar equation of motion 
  involves a second Lie derivative of a brane field as well, 
  namely $\bLn^2\phi$ instead of $\bLn K_{\mu\nu}$ addressed now, it cannot be similarly removed.
  In the string frame in Ref.\ \cite{paper3}, $\bLn K_{\mu\nu}$ and $\bLn^2\phi$ appeared in both directional limits: 
  of the bulk scalar equation and of the bulk tensor equation projected on the brane.
  The former and the trace of the latter constituted a set of linear algebraic equations,
  allowing to determine $\bLn K_{\mu\nu}$ and $\bLn^2\phi$, and to remove them from the directional limits.
  Hence, although no brane equation for the dilaton was obtained, 
  the information contained in the bulk scalar equation 
  was employed in the derivation of the effective brane equations.
} 
\bea
&&\hspace{-1cm}
 \Big[ {\textstyle\frac32} \left( R_{\mu\nu} + K_\mu^\sigma K_{\sigma\nu} - K K_{\mu\nu} \right)
  - {\textstyle\frac38} \, h_{\mu\nu} \left( R + K_\tau^\sigma K_\sigma^\tau - K^2 \right)
  + {\textstyle\frac32} \, E_{\mu\nu}
\nn[2pt]
&&\hspace{1cm}
  - \, {\textstyle\frac13} \, (\pa_\mu\phi) (\pa_\nu\phi)
  + {\textstyle\frac{1}{12}} \, h_{\mu\nu} (\pa\phi)^2 \Big]_{+/-}
  = 0 
 \, ,
 \label{Thh1}
\eea
where the projection of the bulk Weyl tensor on the brane, $E_{\mu\nu}$,
represents the residual dependence on the Lie derivative of the extrinsic curvature, $\bLn K_{\mu\nu}$.

The tensor equation (\ref{Thh1}) can be further rewritten 
in order to obtain a result which resembles the standard Einstein equation.
Clearly, in its current form not only the coefficients of the Ricci tensor $R_{\mu\nu}$ and the Ricci scalar $R$
are non-standard, but also their relative coefficient.
Let us multiply Eq.\ (\ref{Thh1}) by $\frac23$ 
and subtract Eq.\ (\ref{Tnn}) multiplied by $\frac14$ and $h_{\mu\nu}$. 
These steps lead to an Einstein-like tensor equation of motion reading
\bea
&&\hspace{-2.4cm}
 \Big[ R_{\mu\nu} 
  - {\textstyle\frac12} \, h_{\mu\nu} R 
  + K_\mu^\sigma K_{\sigma\nu} 
  - K K_{\mu\nu}
  - {\textstyle\frac12} \, h_{\mu\nu} K^\sigma_\tau K_\sigma^\tau
  + {\textstyle\frac12} \, h_{\mu\nu} K^2 
  - {\textstyle\frac{1}{12}} \, h_{\mu\nu} \, (\bLn\phi)^2
\nn[2pt]
&&\hspace{0.3cm}
  - \, {\textstyle\frac29} \, (\pa_\mu\phi) (\pa_\nu\phi)
  + {\textstyle\frac{5}{36}} \, h_{\mu\nu} \, (\pa\phi)^2 
  + h_{\mu\nu} {\textstyle\frac{1}{2\alpha_1}} V(\phi)
  + E_{\mu\nu} \Big]_{+/-}
  = 0 
 \, . 
 \label{Eeq0}
\eea
However, Eq.\ (\ref{Eeq0}) is just a combination of directional limits (\ref{Thh})-(\ref{W}) of bulk equations of motion
- a dependence on the brane sources (\ref{taus}) is missing.
The junction conditions (\ref{junction1})-(\ref{junction2}) 
relate the jumps in the values of the Lie derivatives of brane fields $h_{\mu\nu}$ and $\phi$,
i.e.\ $\big[ K_{\mu\nu} \big]_\pm$ and $\big[ \bLn\phi \big]_\pm$, respectively,
precisely to the brane sources $\tau_{\mu\nu}$ and $\tau_\phi$ - and thus in principle could introduce such a dependence
into the directional limits (\ref{Thh})-(\ref{W}) of bulk equations, 
and thereby also in their combination - Eq.\ (\ref{Eeq0}).
This possibility has been analyzed in detail in Ref.\ \cite{paper3}.
An analogous discussion can be repeated here, leading to the same result.
Combining the directional limits (\ref{Thh})-(\ref{W}) 
with the junction conditions (\ref{junction1})-(\ref{junction2}) leads to only one non-trivial brane equation
- the consistency condition (\ref{consistency}) on the brane sources (\ref{taus}), reading
\beq
 D_\lambda \! \left( f(\phi) \, \tau_\mu^\lambda \right) 
  = f(\phi) \, \tau_\phi (\pa_\mu\phi)
 \, .
 \label{consistencyA}
\eeq
To be more specific, Eq.\ (\ref{consistencyA}) arises from the difference 
between the `+' and the `-' sides projections (\ref{Thn}), 
when combined with the junction conditions (\ref{junction1})-(\ref{junction2}).

Although Eq.\ (\ref{consistencyA}) does relate the brane fields $h_{\mu\nu}$ and $\phi$
to the brane sources $\tau_{\mu\nu}$ and $\tau_\phi$, its character depends crucially 
on the specific form of the brane matter Lagrangian $\cL_m$.
In particular, apart from more complicated forms of $\cL_m$,\footnote{ 
  E.g.\ involving localized kinetic terms for gravity \cite{k_gravity} 
  or for the scalar field \cite{k_scalar}.} 
Eq.\ (\ref{consistencyA}) does not involve second derivatives (in the brane directions) of brane fields $h_{\mu\nu}$ and $\phi$.
Thus, it is not a dynamical brane equation of motion,
but a consistency condition on the brane sources $\tau_{\mu\nu}$ and $\tau_\phi$,
as is typical of gravity theories \cite{Wald}.

At this point the importance of a $\Z2$ symmetry assumption for the bulk
(with its fixed point at the brane's position) can be shown as follows (and henceforth this symmetry is assumed).
Since the bulk $\Z2$ symmetry relates the quantities at the `+' and the `-' sides of the brane, for example
\beq
 \big[ K_{\mu\nu} \big]_+
  = - \big[ K_{\mu\nu} \big]_-
  = {\textstyle\frac12} \big[ K_{\mu\nu} \big]_\pm
 \quad\textrm{and}\quad
 \big[ \bLn\phi \big]_+
  = - \big[ \bLn\phi \big]_-
  = {\textstyle\frac12} \big[ \bLn\phi \big]_\pm  
 \, ,
\eeq
the junction conditions (\ref{junction1})-(\ref{junction2}) determine now on both sides of the brane
the Lie derivatives $\big[ K_{\mu\nu} \big]_{+/-}$ and $\big[ \bLn\phi \big]_{+/-}$ of brane fields 
(and not only the jumps in their values) in terms of the brane sources.
Let us then substitute the junction conditions into the Einstein-like tensor equation of motion (\ref{Eeq0}),
which results in the effective Einstein-like equation on the brane (\ref{Eeq}), reading
\begin{equation}
 R_{\mu\nu} 
  - {\textstyle\frac12} \, h_{\mu\nu} R 
  = 8\pi \bG(\phi) \, \tau_{\mu\nu}
  - h_{\mu\nu} \bL(\phi)
  + {\textstyle\frac{f^2(\phi)}{4\alpha_1^2}} \pi_{\mu\nu}
  - E_{\mu\nu}
  + {\textstyle\frac29} \, (\pa_\mu\phi) (\pa_\nu\phi)
  - {\textstyle\frac{5}{36}} \, h_{\mu\nu} \, (\pa\phi)^2 
 \, ,
 \label{EeqA}
\end{equation}
where the effective Newton's $\overline{G}$ and cosmological $\overline\Lambda$ constants
are defined in Eqs.\ (\ref{G}) and ({\ref{Lambda}), respectively.
As was already discussed in Section \ref{section_effEinstein},
due to the presence of the bulk Weyl tensor projected on the brane, $E_{\mu\nu}$,
the brane Einstein-like equation (\ref{EeqA}) does not form a closed system. 
Nevertheless, with the bulk $\Z2$ symmetry we can still obtain a genuine equation of motion at the brane
- i.e.\ independent of the bulk's influence.
Specifically, substituting the junction conditions (\ref{junction1})-(\ref{junction2}) 
into the directional limit (\ref{Tnn}) of the bulk tensor equation, yields
\beq
 R
  = {\textstyle\frac{2}{\alpha_1}} V(\phi)
  + {\textstyle\frac13} (\pa\phi)^2 
  + {\textstyle\frac{f^2(\phi)}{4\alpha_1^2}} \left[
    {\textstyle\frac{2\lambda(\phi)}{3f(\phi)}} \tau
    - (\tau\tau) 
    + {\textstyle\frac13} \, \tau^2
    - 3 \, \tau_\phi^2 
    - {\textstyle\frac{6\lambda'(\phi)}{f(\phi)}} \tau_\phi
    + {\textstyle\frac{4\lambda^2(\phi)}{3f^2(\phi)}} 
    - {\textstyle\frac{3\lambda'^2(\phi)}{f^2(\phi)}}  
  \right]
 \, .
 \label{effeq}
\eeq
This is the only `true' effective brane equation of motion,
i.e.\ constraining dynamically the brane fields $h_{\mu\nu}$ and $\phi$
in terms of the brane sources $\tau_{\mu\nu}$ and $\tau_\phi$
- independently of the details of the brane matter Lagrangian $\cL_m$ entering the definitions (\ref{taus}).
However, Eq.\ (\ref{effeq}) is just a single equation.
Nevertheless, it can be of importance e.g.\ in highly-symmetrical setups. 
Moreover, its dependence on the dilaton field can be removed 
with the help of the consistency condition (\ref{consistencyA}).

\section*{Appendix B}
\renewcommand{\theequation}{B.\arabic{equation}}
\setcounter{equation}{0}

In this appendix we will derive a model-independent upper bound 
on the present value $\dot\phi_0$ of the dilaton time derivative,
set by current observational data.
We will follow the main principle of an analogous calculation carried out in a different setup in Ref.\ \cite{dotphi0}.

Let us recall that the spatial derivative of the dilaton field is expected to be not greater than its time derivative (\ref{c1}).
Moreover, the exact value of the upper bound on $\dot\phi_0$ is only of secondary importance
- this analysis aims as showing its smallness.
Therefore, let us assume that our universe can be still to a reasonably good approximation
described by the (modified) Friedmann equations not involving spatial derivatives.

According to the supernovae observations \cite{SNe},  
the universe currently expands at an accelerated rate. 
Hence, the deceleration parameter
\beq
 q 
  \equiv - \frac{\ddot{a}\,a}{\dot{a}^2}
  = - \frac{\ddot{a}}{a} H^{-2}
 \label{q}
\eeq
must be negative,
where $a(t)$ denotes the cosmic scale factor,
and the Hubble constant $H = \frac{\dot{a}}{a}$.
Let us evaluate the effective Einstein-like brane equation (\ref{Eeq})
with the brane-energy momentum tensor of the form of perfect fluid (\ref{fluid})
for the flat Friedmann-Lema\^{\i}tre-Robertson-Walker metric,
and an AdS$_5$ bulk.
Thus, the Friedmann equations read
\bea
&&\hspace{-2.5cm}
 \frac{\dot{a}^2}{a^2} 
  = \frac{8\pi\bG}{3} \rho_m
  - \frac{4\pi\bG}{3} {\textstyle\frac{f(\phi)}{\lambda(\phi)}} \rho_m^2
  + {\textstyle\frac13} \, \overline\Lambda
  + {\textstyle\frac{1}{36}} \, \dot{\phi}^2
 \, ,
 \label{Fried01}
\\[2pt]
&&\hspace{-2.5cm}
 \frac{\ddot{a}}{a} 
  = - \frac{4\pi\bG}{3} \rho_m \left( 1 + 3 \, w_m \right)
  + \frac{4\pi\bG}{3} {\textstyle\frac{f(\phi)}{\lambda(\phi)}} \rho_m^2 \left( 2 + 3 \, w_m \right)
  + {\textstyle\frac13} \, \overline\Lambda
  - {\textstyle\frac{1}{12}} \, \dot{\phi}^2
 \, ,
 \label{Fried02}
\eea
where $\rho_m$ stands in principle for any kind of matter, 
and the coefficient of the $\rho_m^2$ term was rewritten 
using the definition of the effective brane Newton's constant (\ref{G}).
Since the present radiation contribution to the content of the universe is negligible, 
we set $w_m=0$ henceforth.

The contributions to the Friedmann equations (\ref{Fried01})-(\ref{Fried02})
due to the effective brane cosmological constant (\ref{Lambda})
and the kinetic energy density of the dilaton, 
can be rewritten in a form similar to the matter contribution $\rho_m$.
Consequently, the Friedmann equations (\ref{Fried01})-(\ref{Fried02}) become
\bea
&&
 \frac{\dot{a}^2}{a^2} 
  = \frac{8\pi\bG}{3} \sum_{i=m,\overline\Lambda,\phi} \rho_i
  - \frac{4\pi\bG}{3} {\textstyle\frac{f(\phi)}{\lambda(\phi)}} \rho_m^2
 \, ,
 \label{Fried1}
\nn[2pt]
&&
 \frac{\ddot{a}}{a} 
  = - \frac{4\pi\bG}{3} \sum_{i=m,\overline\Lambda,\phi} \rho_i \left( 1 + 3 \, w_i \right)
  + \frac{8\pi\bG}{3} {\textstyle\frac{f(\phi)}{\lambda(\phi)}} \rho_m^2
 \, ,
 \label{Fried2}
\eea
where $w_{\overline\Lambda}=-1$ and $w_\phi = \frac53$.\footnote{
  In the standard cosmology, for a scalar field we have $w \in [-1,1]$.
  Nevertheless, the value of $w_\phi = \frac53$ can be easily understood as follows. 
  In Ref. \cite{dotphi0}, the relative coefficient of dilaton kinetic terms 
  $(\pa_\mu\phi)(\pa_\nu\phi)$ and $h_{\mu\nu}(\pa\phi)^2$ in the gravitational equation is of the standard value $\frac12$.
  However, the theory in Ref. \cite{dotphi0} is 4-dimensional, whereas in the present work,
  the original value of $\frac12$ of this relative coefficient in the bulk tensor equation (\ref{tensor0}) 
  changes during the derivation of the effective brane equations. 
  To be more specific, the coefficient of $h_{\mu\nu}(\pa\phi)^2$ is modified twice: 
  when the relative coefficient of $R_{\mu\nu}$ and $h_{\mu\nu}R$ terms is corrected, 
  and when the dependence on the Lie derivative of the extrinsic curvature, $\bLn K_{\mu\nu}$, is removed. 
  Therefore the non-standard value of $w_\phi = \frac53$
  is understood as a consequence of the higher-dimensional nature of the model.}
The contributions involving $\rho_m^2$ could have played an important role at the early stages of the universe's evolution, 
but they are not being observed at later times, and thus can be neglected here.
Therefore, the deceleration parameter (\ref{q}) reads
\beq
 q 
  = {\textstyle\frac12} \sum_{i=m,\overline\Lambda,\phi} \Omega_i \left( 1 + 3 \, w_i \right)
 \, .
\eeq
Taking into account the respective values of the equation of state parameters $w_i$ 
for the universe energy density components $\Omega_m$, $\Omega_{\overline\Lambda}$, and $\Omega_\phi$,
for the flat universe with $\Omega_m + \Omega_{\overline\Lambda} + \Omega_\phi = 1$ we get
\beq
 |\dot{\phi}|
  = 3 H \left( 1 + q - {\textstyle\frac32} \, \Omega_m \right)^{1/2}
 \, ,
\eeq
as $\Omega_\phi = {\textstyle\frac{1}{36H^2}} \, \dot{\phi}^2$ follows from Eq.\ (\ref{Fried01}).\footnote{
  It is assumed that the critical density can be formulated as $\rho_c = \frac{3H^2}{8\pi\bG}$ approximately. }
Substituting the present epoch's values: $q_0<0$,
$\Omega_{m0}>0.25$ and $H_0 \simeq 72 \frac{\textrm{km}}{\textrm{s}\cdot\textrm{Mpc}}$ \cite{PDG}, 
we obtain an upper bound on the time derivative of the dilaton, reading
\beq
 |\dot{\phi}_0| 
  \lesssim 2.4 \, H_0
  \simeq 1.8 \, \big( 10^{10} \, \textrm{yr} \big)^{-1}
 \, .
\eeq



\begin{thebibliography}{99}


\bibitem{lowenergy}
  B.~Zwiebach,
  Phys.\ Lett.\  B {\bf 156} (1985) 315;
  D.G.~Boulware and S.~Deser,
  Phys.\ Rev.\ Lett.\  {\bf 55} (1985) 2656;
  Phys.\ Lett.\  B {\bf 175} (1986) 409;
  R.~R.~Metsaev and A.~A.~Tseytlin,
  Phys.\ Lett.\  B {\bf 191} (1987) 354;
  D.~J.~Gross and J.~H.~Sloan,
  Nucl.\ Phys.\  B {\bf 291} (1987) 41.
\bibitem{paper2}
  D.~Konikowska and M.~Olechowski,
  Phys.\ Rev.\  D {\bf 76} (2007) 124020
  [0704.1234[hep-th]].
\bibitem{meissner}
  K.~A.~Meissner,
  Phys.\ Lett.\ B\ {\bf 392} (1997) 298
  [hep-th/9610131].
\bibitem{brane}
  K.~Akama,
  Lect.\ Notes Phys.\  {\bf 176} (1982) 267
  [hep-th/0001113]; 
  V.~A.~Rubakov and M.~E.~Shaposhnikov,
  Phys.\ Lett.\  B {\bf 125} (1983) 136.
\bibitem{HoWi}
  P.~Horava and E.~Witten,
  Nucl.\ Phys.\  B {\bf 460} (1996) 506
  [hep-th/9510209];
  Nucl.\ Phys.\  B {\bf 475} (1996) 94
  [hep-th/9603142].
\bibitem{review}
  for a review see e.g.\
  R.~Maartens and K.~Koyama,
  Living Rev.\ Rel.\ \ {\bf 13} (2010) 5
  [1004.3962[hep-th]].
\bibitem{paper3}
  D.~Konikowska and M.~Olechowski,
  Class.\ Quant.\ Grav.\ \ {\bf 27} (2010) 145015
  [0908.1052[hep-th]].
\bibitem{SaShMa}
  T.~Shiromizu, K.~i.~Maeda and M.~Sasaki,
  Phys.\ Rev.\  D {\bf 62} (2000) 024012
  [gr-qc/9910076].
\bibitem{MaWa}
  K.~i.~Maeda and D.~Wands,
  Phys.\ Rev.\  D {\bf 62} (2000) 124009
  [hep-th/0008188].
\bibitem{MeBa}
  A.~Mennim and R.~A.~Battye,
  Class.\ Quant.\ Grav.\  {\bf 18} (2001) 2171
  [hep-th/0008192].
\bibitem{conformal}
  V.~Faraoni, E.~Gunzig and P.~Nardone,
  Fund.\ Cosmic Phys.\  {\bf 20} (1999) 121
  [gr-qc/9811047].
\bibitem{ads/cft}
  J.~M.~Maldacena,
  Adv.\ Theor.\ Math.\ Phys.\  {\bf 2} (1998) 231
   [Int.\ J.\ Theor.\ Phys.\  {\bf 38} (1999) 1113]
  [hep-th/9711200];
  O.~Aharony, S.~S.~Gubser, J.~M.~Maldacena, H.~Ooguri and Y.~Oz,
  Phys.\ Rept.\  {\bf 323} (2000) 183
  [hep-th/9905111].
\bibitem{RS}
  L.~Randall and R.~Sundrum,
  Phys.\ Rev.\ Lett.\  {\bf 83} (1999) 3370
  [hep-ph/9905221];
  L.~Randall and R.~Sundrum,
  Phys.\ Rev.\ Lett.\  {\bf 83} (1999) 4690
  [hep-th/9906064].
\bibitem{PDG}
  K.~Nakamura {\it et al.}  [Particle Data Group Collaboration],
  J.\ Phys.\ G G {\bf 37} (2010) 075021.
\bibitem{Uz}
  J.~-P.~Uzan,
  Living Rev.\ Rel.\ \ {\bf 14} (2011) 2
  [1009.5514[astro-ph.CO]].
\bibitem{SDSS}
  D.~G.~York {\it et al.} [SDSS Collaboration],
  Astron.\ J.\ \ {\bf 120} (2000) 1579
  [astro-ph/0006396];
  H.~Aihara {\it et al.} [SDSS Collaboration],
  Astrophys.\ J.\ Suppl.\ \ {\bf 193} (2011) 29
   [Erratum-ibid.\ \ {\bf 195} (2011) 26]
  [1101.1559[astro-ph.IM]].
\bibitem{2dFGRS}
  M.~Colless {\it et al.} [The 2DFGRS Collaboration],
  Mon.\ Not.\ Roy.\ Astron.\ Soc.\ \ {\bf 328} (2001) 1039
  [astro-ph/0106498].
  M.~Colless, B.~A.~Peterson, C.~Jackson, J.~A.~Peacock, S.~Cole, P.~Norberg, I.~K.~Baldry and C.~M.~Baugh {\it et al.},
  [astro-ph/0306581].
\bibitem{book}
  H.~Mo, F.~van den Bosch and S.~White
  ``Galaxy Formation and Evolution'',
  Cambridge University Press 2010.
\bibitem{plot}
  see e.g.\ Fig.\ 3 in E.~A.~Kazin {\it et al.}  [SDSS Collaboration],
  Astrophys.\ J.\  {\bf 710} (2010) 1444
  [0908.2598[astro-ph.CO]].
\bibitem{k_gravity}
  G.~R.~Dvali, G.~Gabadadze and M.~Porrati,
  Phys.\ Lett.\  B {\bf 485} (2000) 208
  [hep-th/0005016];
  M.~S.~Carena, A.~Delgado, J.~D.~Lykken, S.~Pokorski, M.~Quiros and C.~E.~M.~Wagner,
  Nucl.\ Phys.\  B {\bf 609} (2001) 499
  [hep-ph/0102172];
  G.~R.~Dvali, G.~Gabadadze, M.~Kolanovic and F.~Nitti,
  Phys.\ Rev.\  D {\bf 64} (2001) 084004
  [hep-ph/0102216];
  R.~Bao, M.~S.~Carena, J.~Lykken, M.~Park and J.~Santiago,
  Phys.\ Rev.\  D {\bf 73} (2006) 064026
  [hep-th/0511266].
\bibitem{k_scalar}
  F.~del Aguila, M.~Perez-Victoria and J.~Santiago,
  JHEP {\bf 0302} (2003) 051
  [hep-th/0302023];
  C.~Csaki, J.~Hubisz and P.~Meade,
  [hep-ph/0510275];
  M.~Olechowski,
  Phys.\ Rev.\  D {\bf 78} (2008) 084036
  [0801.1605[hep-th]].
\bibitem{Wald}
  see for example: R.\ Wald ``General Relativity'' 
  The University of Chicago Press 1984.
\bibitem{dotphi0}
  T.~Damour, F.~Piazza and G.~Veneziano,
  Phys.\ Rev.\ D {\bf 66} (2002) 046007
  [hep-th/0205111].
\bibitem{SNe}
  A.~G.~Riess {\it et al.}  [Supernova Search Team Collaboration],
  Astron.\ J.\  {\bf 116} (1998) 1009
  [astro-ph/9805201].
\end{thebibliography}
\end{document}